\documentstyle[prd,aps,preprint]{revtex}
\tightenlines
\begin{document}
\draft

\def\beq{\begin{equation}}
\def\eeq{\end{equation}}
\def\bea{\begin{eqnarray}}
\def\eea{\end{eqnarray}}
\def\simlt{\stackrel{<}{{}_\sim}}
\def\simgt{\stackrel{>}{{}_\sim}}
\input epsf

\title{What if Dark matter is Bosonic and self-interacting}
\author{Antonio Riotto$^{(1,2)}$ and Igor Tkachev$^{(3,4)}$}

\address{$^{(1)}${\it Scuola Normale Superiore, Piazza dei Cavalieri 7, 
I-56126 Pisa, Italy}}
\address{$^{(2)}${\it INFN, Sezione di Pisa, 
I-56127 Pisa, Italy}}

\address{$^{(3)}${\it Institute for Theoretical Physics, ETH-Hoenggerberg,
        CH-8093 Zurich, Switzerland }}
\address{$^{(4)}${\it Institute for Nuclear Research of the 
Academy of Sciences of Russia, Moscow 117312, Russia }}

\def\be{\begin{equation}}
\def\ee{\end{equation}}
\def\bea{\begin{eqnarray}}
\def\eea{\end{eqnarray}}
\def\la{\mathrel{\mathpalette\fun <}}
\def\ga{\mathrel{\mathpalette\fun >}}
\def\fun#1#2{\lower3.6pt\vbox{\baselineskip0pt\lineskip.9pt
        \ialign{$\mathsurround=0pt#1\hfill##\hfil$\crcr#2\crcr\sim\crcr}}}
\def\rhoeq{{{\rho_{\rm eq}}}}
\def\rhodm{{{\rho_{\rm DM}}}}
\def\Teq{{{T_{\rm eq}}}}
\def\aeq{{{a_{\rm eq}}}}
\def\bra{{{\bar{\rho}_a}}}

\maketitle
\begin{abstract}
Recently the problem of singular galactic cores and over-abundant formation of
dwarf galaxies, inherent to the standard cold dark matter model, had
attracted  a great deal of attention. One scenario which may be free of these
problems invokes a self-interacting
Bose-field. We find the limiting  core density in this model due
to the  self-annihilation of the scalar 
field into its own relativistic quanta. 
The limiting density may correspond
to the observable one if there is only one dark matter component. 
Alternatively, there may be  more than one dark matter species and the 
annihilation of one species 
may be  very efficient with subsequent expansion of the other, thus
avoiding the problem of singular cores.
\end{abstract}

\def\simlt{\stackrel{<}{{}_\sim}}
\def\simgt{\stackrel{>}{{}_\sim}}

\newpage

{\it Introduction.}~~
The nature of the dark component which constitutes most of  the matter
 in the Universe remains unknown. It is usually assumed that 
the dark matter (DM)  is cold and non-interacting. Adding  a  non-zero
cosmological constant to the non-baryonic dark matter gives rise to the
so-called  $\Lambda$CDM model. In spite of its great success in
reproducing the observed Universe at large scales, 
in the recent few  years it has become  more and more evident
 that the  $\Lambda$CDM model suffers from predicting
too much  power of density perturbations on small scales.
First, the central density distributions in dwarf and low surface
brightness galaxies seem to have  finite cores in contrast \cite{fc} with the
singular profiles predicted by N-body simulations \cite{nb}
(see however Ref.\cite{krav}). Secondly, high resolution
numerical experiments predict that galaxies as our own
should contain several thousands clumps of dark matter  of the size
of a dwarf galaxy \cite{clumps}. However,  only a few  of these are observed.

Several solutions to this problem have been  suggested in the literature:
{\it i)} the  dark matter is not cold, but warm \cite{warm};
{\it ii)} the dark matter is interactive \cite{interacts};
{\it iii)} the dark matter is warm and interactive \cite{hs};
{\it iv)} the power spectrum of $\Lambda$CDM has sharp drop
on subgalactic scales \cite{drop};
{\it v)} the dark matter is in the form of a self-interacting 
scalar field \cite{scf}.
All these ideas are not new and have been previously considered in different
applications. For instance,  warm dark matter was considered 
in \cite{wdm}, 
self-interacting dark matter was studied in Ref. \cite{intdm},
while the scalar field model of Ref. \cite{scf} was 
considered in Ref. \cite{it86}. 

Clearly all these models require an  appropriate degree of fine-tuning
in their parameters ({\it e.g.}  mass and self-couplings).
For example, the dark matter component  of Ref. \cite{interacts} 
should interact
sufficiently strongly to allow the dark matter in the halo 
to start relaxing on time-scale of the age of the Universe, 
but the relaxation should  
not be too fast and already completed otherwise  a gravothermal instability
will inevitably occur \cite{grinst}, leading to the formation
of  singular profiles.
This is not  a problem within  the scalar field model where
the pressure which supports the core is generated by the self-repulsion
of dark matter particles. However, for the halo  dominated by the scalar field 
the radius of its core is only a function of the field mass and 
self-coupling \cite{it86,scf} and  therefore they should obey 
certain relation to fit observations.

In this paper we will study  further constraints on the parameters
of scalar field  model with mass $m$ and self-coupling $\lambda$ 
(with interaction potential defined as $\lambda \phi^4/4$) and give
a testable prediction for the core radius and core density of the
gravitational bound object composed by scalar quanta. In particular, we
analyze constraints which are due to inevitable 
self-annihilation of the self-interacting scalar 
field without conserved quantum numbers into its own relativistic quanta.

{\it Self-interacting scalar field and gravitationally bound clumps.}~~ 
It is assumed that the scalar field is in the state
of coherent oscillations, just like the familiar axion in cosmological
setting,
and that the gravitationally bound object is formed out of the
field clump. Of the prime interest for comparison with observations is the
case when self-gravity of the field dominates (generalization for the case of
the scalar field in the external gravitational well, created e.g. by baryonic
component, is straightforward). The parameters of such configuration can be
estimated analyzing a  simple equation of 
``hydrostatic equilibrium'' in non-relativistic limit \cite{it86}
\begin{eqnarray}
{dP(r) \over dr} &=& -{\rho (r) M(r) \over M_{\rm Pl}^2 r^2} \, , 
\\
{dM(r) \over dr} &=& 4\pi r^2 \rho (r) \, .
\label{he}
\end{eqnarray}
The pressure $P(r)$ and the density $\rho(r)$ 
have  to be understood here as quantities averaged over the period of field 
oscillation, which gives for the equation of state \cite{it86,scf}
$P \approx \lambda\rho^2/3m^4 $. These equations have a simple solution
$\rho(r) = \rho(0) {\rm sin}(x)/x$, where $x \equiv r/r_c$ and 
the core radius satisfies\footnote{This definition of $r_c$ coincides with
Ref. \cite{it86} but is $\sqrt{6}$ smaller compared to Ref.   \cite{scf},
because in \cite{scf} the $r_c$ was defined via $\rho(r_c) =0$.}
\begin{equation}
\frac{m^4}{\lambda} \approx \frac{M_{\rm Pl}^2 }{6\pi r_c^2} \, ,
\end{equation}
or 
\begin{equation}
r_c \sim 10^2\, \sqrt{\lambda}\, ({\rm eV}/m)^2 \, {\rm kpc} \, .
\label{rc}
\end{equation}
Note that the density in the core, $\rho(0)$, is unrelated to the
core radius, but the density is limited. Equation of state changes
to relativistic, $P = \rho /3$ when 
$\phi_0 > m/\sqrt{\lambda}$, where $\phi_0$ is the
amplitude of field oscillations, and the object became unstable \cite{it86}.
This corresponds to the maximum density, $\rho = m^2 \phi_0^2$, 
\begin{equation}
\rho_{\rm max} \sim m^4/\lambda \, .
\label{rhomax1}
\end{equation}
This density corresponds to the object which is close to the
limit of black hole formation. Since  
this is the only restriction on the density if the field
has conserved quantum numbers (which corresponds to the complex scalar field),
the model with conserved particle number is unappealing
(the excess dark matter cusps may Bose-condense into black holes
in the centers of galaxies \cite{it91}, but this can not be a solution for all
systems). Therefore,
here we consider the model where the dark matter particle number
is not conserved. For example, 
the real scalar field can decay or
self-annihilate. The corresponding constraints on the density 
arising from self-annihilation, which are
inevitable because of the non-zero $\lambda$,
will be considered below. Let us first though  consider under
which conditions the coherent field configuration  forms.

{\it Condensation.}
If the self-coupling is not very large, during the collapse
of spatially large configurations
the coherence of the field will be destroyed even if the
configuration develops from the small overdensity in initially
homogeneous oscillating field. Namely, after virialization 
at each spatial point there
will be many streams of particles, each with different vector of velocity.
In such situation  the  analysis above  and Eq. (\ref{rc}) are
not applicable. If velocities dominate, 
the virialized configuration will behave in the gravitational
field as non-interacting usual dark matter. However, with time
the Bose-condensation will occur \cite{it91}. This will happen even with
very small values of the self-coupling since the  scattering is Bose-enhanced
if the phase-space density of particles is larger than unity,
which will be always the case with parameters satisfying
Eq. (\ref{rc}). If the relaxation time for the  Bose-condensation
is smaller than the age of the Universe, we may apply Eq. (\ref{rc})
for the final configuration.

Let us consider now the  relaxation in virialized clumps due to
the scattering process $2\phi \rightarrow 2\phi$ following Refs.
\cite{it91,st}.
The inverse relaxation time is
$t_R^{-1} \sim {(1+n)}  \sigma \rho v_{\rm e} m^{-1}$, where
$\sigma$ is the corresponding cross section, and $v_{\rm e}$ is 
the characteristic velocity which characterizes the depth of the 
gravitational well. We are taking into account the possibility
that mean phase-space 
density of $\phi$ particles, ${n}$, can be large, which accounts
for the factor $(1+n)$ in the expression for the relaxation time.
For particles bound
in a gravitational well, it is convenient to rewrite this
expression in the form \cite{it91}
\begin{equation}
t_R^{-1}  \sim  \frac{\lambda^{2}\rho v_{\rm e}} {m^{3}}
\left[ 1 + \frac{\rho}{m^{4}v_{\rm e}^{3}} \right]\, .
\label{rt}
\end{equation}
Let us consider first the case $n \gg 1$. The relaxation time in 
the clump is smaller than a given time $\tau$
if 
\begin{equation}
\rho >  \frac{v_e}{\sqrt{m\tau}}\frac{m^4}{\lambda} \, .
\label{rhocond}
\end{equation}
With $\rho = 0.02 M_{\odot}/{\rm pc}^3 \approx 5\times 10^{-6}$ 
${\rm eV}^4$ and $v_e = 100$ km s$^{-1}$, 
which approximately corresponds to the parameters of the 
cores of dwarf galaxies, 
the relaxation time
will be smaller than the age of the Universe if 
\begin{equation}
\lambda >  10^{-15} \, (m/{\rm eV})^{7/2}.
\label{rl}
\end{equation}
If this condition  is satisfied, the Bose condensate in the center
of gravitational well should form \cite{it91}.

In addition to the condensation due to self-interaction, there is
a process of purely gravitational relaxation, also with subsequent formation
of the coherent field configuration, see Refs. \cite{gravcool}. However, this
should be 
efficient only if density inhomogeneities are large (e.g. during initial stage
of the collapse),
while we are interested in continuing condensation in the already 
virialized clump as well. In addition, in all range of parameters 
relevant for the present discussion, the condensation solely due to 
self-coupling is efficient. 

{\it Self-annihilation of the condensate.}
Besides taking into account the  specific Bose-enhancement
during the process of collisional relaxation, one has also to 
consider the decay of the  condensate. This is 
 also peculiar and may correspond to a ``laser'' effect \cite{it86,it87}.
In other words, the Bose-stimulation of relevant processes has to be taken into
account.

 In such a  case it is important to know up to which distances the
decay  or annihilation products stay in the resonance with each other. 
If the initial configuration is not condensed, but can be described by some
distribution of (free) particles over momenta, one can employ the Boltzmann
kinetic 
approach \cite{it87}. In a stationary state (created, say, by the process of
violent relaxation) the distribution of particles 
in the phase space is a function of integrals of  motion like 
energy. This does not mean though that one can neglect the 
red-shift in the collision integral for particles which are at the same energy
level. This is  because
particles are moving. However, there is enough phase-space available
for the decay products to stay in the resonance if the distribution 
of ``parent'' particles is isotropic
\cite{it87}. With the average momentum of the distribution of the 
parent particles
going to zero (condensation), the collision integral start to diverge. The
Boltzmann approach 
breaks down and one can employ \cite{it86} the formalism of particle
creation by time dependent classical background.

For the condensed state in  equilibrium in the gravitational well we
have the following peculiarity: the red-shift cancels out and 
the  decay products stay in
the resonance throughout the whole configuration. First, in the condensed state
particles are not moving. Second, in the state of hydrostatic equilibrium
the gravitational energy and internal interaction energy are tuned precisely
in the way to cancel the gravitational 
redshift. This can be understood in several ways.
Indeed, in hydrostatic equilibrium, the total mass of configuration $M$
 (which
includes all forms of energy) takes its minimum possible value under
perturbations of internal structure. Some small amount of particles can be
moved around resulting in  $\delta M =0 $. As a Gedanken experiment one can
think to  move around
particles which are going to decay. The condition $\delta M =0 $ means that at
infinity the products of decay will have the same energy. 
Therefore, they stay in
resonance on WKB trajectories labeled by quantum numbers at infinity.
In other words, from the point of view of local observers, in hydrostatic
equilibrium $\mu \sqrt{g_{00}}$ = const, 
where $\mu$ is chemical potential, see e.g. \cite{ZN}. 
Compare this with the energy of 
free moving particle as  measured by the local observer, $\omega \sqrt{g_{00}}$
= const. Therefore, $\mu/\omega =$ const and the decay process stays in the
resonance.

Let us make this explicit for the self-annihilation of the 
Bose-condensed field in the metric $ds^2 = fdt^2 - f^{-1} \delta_{ij}dx^idx^j$,
where $f = 1 + 2\Phi$ and $\Phi$ is Newtonian gravitational potential.
Condensed state can be described by the field configuration of
the form $\phi = \phi(t)\psi(\bf{x})$, where $\phi(t)$ is periodic function of
time which to the first approximation is $\phi(t) = \phi_0 \sin \omega t$,
with $\omega$ close to $m$.
Let us describe quanta which propagate in this metric as WKB plane waves,
$\delta \phi \propto g_p(t) e^{ip_j x^j}$, where $p_j= p_j(\bf{x})$ 
are spatial momenta which satisfy 
$g^{\mu\nu} p_\mu p_\nu = m_{\rm eff}^2$ or $p_0^2 = f^2 |{\bf p}|^2 + f
m_{\rm eff}^2 = $ const. 
Here $m_{\rm eff}$ includes contribution to the particle mass 
from the interaction 
with the condensate, $m_{\rm eff}^2 = m^2 + 3\lambda\phi_0^2\psi^2/2$.
Mode functions satisfy the equation 
\begin{equation}
\delta\ddot{\phi} - f^2 \nabla^2 \delta \phi + f(m_0^2 + 3\lambda \phi) \delta
\phi = 0 \, .
\label{mfe}
\end{equation}
Therefore the equation for $g_p(t)$ is
\begin{equation}
\ddot{g_p} + p_0^2 g_p - 2\omega^2(1+2\Phi)\psi^2  q \cos(2\omega t)g_p = 0 \,
, 
\label{mfeg}
\end{equation}
where 
$q \equiv 3\lambda\phi_0^2/4\omega^2$ 
is the resonance parameter.
We see that change in the gravitational parameter $\Phi$ with the distance
gives insignificant 
change in the ``effective'' resonance parameter. More important is the change
in $\psi^2(\bf{x})$, but most of the
modes with the same value of $p_0$ can stay in the resonance up to the
distances comparable to the core size. The important fact which leads to this
conclusion is that $\omega =$ const. 

The problem reduces to particle creation by time dependent homogeneous 
classical background and is very well studied, see
{\it e.g.} \cite{gmm}. We briefly review the results below.
The number density of the created particles grows exponentially with time,
$n_k = \exp (\mu_k t)$ where the characteristic exponent is 
positive and non-zero
in narrow resonance bands and its numerical value is model dependent, being
a function of the coupling constants of the theory. There can be several
channels of decay, e.g. $\phi$ can be coupled to photons leading to peculiar
cosmic maser effect discussed in \cite{it86} which by itself puts constraints
on the strength of the coupling to the electromagnetic field.  
The strength of the coupling $f_\phi$  to the electromagnetic field
(as well as to any 
other possible but hypothetical field) is unknown and might be even zero.
However, we cannot disregard   the decays of   the condensate
into its own quanta because of the self-coupling $\lambda$. 

Let us 
consider the consequences of the condensate decays.
The rate of particle production as a function of particle momenta $k$
is determined by the growth rate of unstable solutions of the
Mathieu equation for the corresponding mode functions
\begin{equation}
\ddot{g}_k + [A -2q \cos (2\tau)] g_k =0 \, ,
\label{me}
\end{equation}
and at the center of the $N$-th instability band the parameter $\mu_N$
is given by
\begin{equation}
\mu_N = \frac{m}{2N}\frac{q^N}{(2^{N-1}(N-1)!)^2}  \, .
\label{mu}
\end{equation}
The coupling of $\phi$ to the electromagnetic field gives \cite{it86}
$A = 4k^2/m^2$ and $q = 2k\phi_0/mf_\phi$. The
decay of $\phi$ into two photons is saturated in the first
instability band which is centered at
$A=1$. In such a case,  $\mu_1 = qm/2$ and the products of the
 decay have momentum $k = m/2$,
with a width $\delta k \approx \mu$.

For the self-annihilation $4\phi \rightarrow 2\phi$,  as we've seen one finds
$A = (k^2+m^2)/m^2 + 2q$ and 
\begin{equation}
q = 3\lambda\phi_0^2/4m^2 \, .
\label{q}
\end{equation}
The self-decay of $\phi$ occurs in the second instability band 
centered at $A=4$ with
$\mu_2 = q^2 m/16$, momentum  $k = \sqrt{3}m$ and  
 width $\delta k \approx \mu$.
Clearly, the products of the self-annihilation
are ultra-relativistic and easily escape the gravitational well.

The rate of decay $\mu$ is a function of the amplitude of the field
oscillations $\phi_0$ and therefore  is a function of the energy
density in the core: $\rho = m^2\phi_0^2$. Notice that
the exponential growth  of the  particle 
number in the resonance bands is due to  the  Bose-statistics: 
already created particles stimulate
production of new quanta. However, particles which leave gravitational well 
do not participate in the stimulation process.
Therefore, the exponential growth occur only if $D \equiv \mu r_c > 1$, 
where $r_c$ is  the core radius of the field configuration.
If $D \gg 1$ initially, the number density of decay quanta will 
grow exponentially in time in the region of the core \cite{it87} 
because in each decay process two identical particles are produced
and  travel in opposite directions. In a sense, this system is equivalent 
to the inversely 
populated laser medium placed between reflecting mirrors. 
The resulting explosion
will reduce the density in the core below the level at which $D \simeq 1$. 
If the core was  growing gradually, starting from the small density, 
which is likely the case in astrophysical situation, the density will just 
stop growing at the condition $D \approx 1$ even if the infall of 
particles continues due to condensation from surrounding 
non-relativistic ``gas''. In this regime the
luminosity in relativistic particles will be equal to the rate of 
Bose-condensation.

Let find the maximum core density corresponding to the condition
$D \sim 1$ in the case of self-annihilation. The condition $q^2 m r_c\sim 1$ 
with $r_c$ defined in Eq. (\ref{rc}) and $q$ defined in Eq. (\ref{q})
gives\footnote{Numerically, the assumption of perfect hydrostatic equilibrium
turns out to be not vitally
important here. Had we used instead of $r_c$ the distance over which
the gravitational redshift equals to the width of the resonance band, we would
have obtained exactly the same relation, but with power 1/2 replaced by the
power 2/5 in the prefactor of $ m^4/\lambda$.}
\begin{equation}
\rho_{\rm max} \sim \left(\frac{m}{M_{p}\sqrt{\lambda}}\right)^{1/2}\,
\frac{m^4}{\lambda} \sim \left(\frac{1}{mr_c}\right)^{1/2}\,
\frac{m^4}{\lambda} \, .
\label{rhomax2}
\end{equation}
Interestingly, there are indications that the halo central density
is nearly independent of the mass from the galactic to the galaxy cluster
scales, with average value of around 
$\rho = 0.02 M_{\odot}/{\rm pc}^3$ \cite{firma}.
With this value of density, Eq. (\ref{rhomax2}) gives
\begin{equation}
\lambda \approx 10^{-8} \, (m/{\rm eV})^{7/2} \, .
\label{lm}
\end{equation}
Comparing this with the condition (\ref{rl}), we see
that the Bose-condensation is efficient indeed in this parameter range.

Let us finally estimate the maximum core density which
corresponds to non-stimulated self-annihilation ($D < 1$,
and the rate of Bose-condensation is small).
The core density will stop changing effectively when 
the rate of annihilation will became comparable to the
age of the Universe, $t_0$. We find
\begin{equation}
\rho_{\rm c} \sim \frac{1}{(mt_0)^{1/3}}\frac{m^4}{\lambda} \, .
\label{rhoc}
\end{equation}
\frenchspacing 

We see that not only the core radius, but also the core density  may be
uniquely defined in terms of  the mass and the coupling of the scalar
field. 
This provides a  severe test
of the model. The core radius 
or the maximum density (\ref{rhomax2}) will be changed somewhat, 
if the field is embedded
in an  external gravitational well (created by baryonic matter),
but this will not alter the required parameter range significantly.

However, the picture may be more complicated and perhaps more interesting
with core radius  (\ref{rc}) and limiting density (\ref{rhomax2})
not related to the observed characteristics of the dark halos.
Assume that initially the
field $\phi$ contributes to the most of the  dark component and  that  the
parameters
are such that the Bose-condensation is efficient on a  time
scale shorter than the age of the Universe. Finally, suppose  that in the core
of the halo which forms the condition $D \gg 1$ is satisfied. 
This is a modification of the scenario considered in
Refs. \cite{it86,it87,it91} where the possibility of electromagnetic
radiation was advocated. Here annihilation of the field $\phi$ into
itself will do the job, but we require that the  Bose-condensation
with subsequent inflow into the core and decays
are efficient enough to re-process the major fraction
of the dark halo into relativistic particles.
The baryonic core (and any other dark component) will then
expand after loosing part of gravitating central mass. The problem
of singular cores may be avoided in this way in a wider range of parameters.

Needless to say, our proposal of annihilating dark matter, as a solution of CDM
problems on small scales, is not
limited to a self-annihilating scalar field, but applies as well to
other possible forms of dark matter particles which may annihilate into
different species. 

We are grateful to V. A. Berezin for useful discussions.


\begin{thebibliography}{99}
\frenchspacing 
\bibitem{fc}
B. Moore, Nature 370 (1994) 629;
R. Flores and J.R.Primack, Ap. J. 427 (1994) L1.
\bibitem{nb}
J. Dubinski and R. Carlberg, Ap. J. 378 (1991) 496;
J.F. Navarro, C.S. Frenk and S.D.M. White, Ap. J. 462 (1996) 563.
\bibitem{krav}
A.V. Kravtsov, A.A. Klypin, J.S. Bullock and J.R. Primack,
astro-ph/9708176.
\bibitem{clumps}
B. Moore, S. Ghigna,  F. Governato, G. Lake, T. Quinn,
J. Stadel, and P. Tozzi, Ap. J. 524 (1999) L19;
S. Ghigna, B. Moore, F. Governato, G. Lake, T. Quinn
and J. Stadel, astro-ph/9910166.
\bibitem{warm}
R.~Schaefer and J.~Silk, Ap. J. 332 (1988) 1;
J.~Sommer-Larsen and A.~Dolgov, astro-ph/9912166;
C.~J.~Hogan and J.~J.~Dalcanton, astro-ph/0002330.
\bibitem{interacts}
D.~N.~Spergel and P.~J.~Steinhardt, 
astro-ph/9909386.
\bibitem{hs} S. Hannestad and R. Scherrer, astro-ph/0003046.
\bibitem{drop}
M.~Kamionkowski and A.~R.~Liddle,
astro-ph/9911103.
\bibitem{scf} P.~J.~E.~Peebles, astro-ph/0002495;
J. Goodman, astro-ph/0003018.
\bibitem{wdm} J.R. Bond, G. Efstathiou and J. Silk,
Phys. Rev. Lett. 45 (1980) 1980;
J. Bardeen et al. Ap. J. 304 (1986) 15;
G. Blumenthal, H. Pagels and J. R. Primack,  Nature 299 (1982) 37;
A. Melott and D. N. Schramm, Ap. J. 298 (1985) 1. 
\bibitem{intdm}
C. Dover, T, Gaisser, and G. Steigman, Phys. Rev. Lett. 42  (1979)
1117;
S. Wolfram, Phys. Lett. B 82 (1979) 65;
G.~Raffelt and J.~Silk, Phys. Lett. B 192 (1987) 65;
G. Starkman et al., Phys. Rev. D 41 (1990) 2388;
F.~Atrio-Barandela and S.~Davidson, Phys. Rev. D 55 (1997) 5886;
E.~D.~Carlson, M.~E.~Machanek and L.~J.~Hall, Ap. J. 398 (1992) 43;
M.~E.~Machacek, Ap. J. 431 (1994) 41;
A.~A.~de Laix, R.~J.~Scherrer and R.~K.~Schaefer, Ap. J. 452 (1995) 495;
A. D. Dolgov, S. Pastor and J.W.F. Valle, astro-ph/9506011;
R. N. Mohapatra and V. L. Teplitz, Phys. Rev. Lett. 81 (1998) 3057;
R. N. Mohapatra and V. L. Teplitz, astro-ph/0001362.
\bibitem{it86}
I. I. Tkachev, Sov. Astron. Lett. 12 (1986) 305;
\bibitem{grinst} V. A. Antonov, Vestn. Leningr. Univ., Mat. Mekh. Astron., 
135-146, 1962; CERN-TRANS 77-05.
\bibitem{it91}
I. I. Tkachev, Phys. Lett. B 261 (1991) 289.
\bibitem{st}
D. V. Semikoz and I. I. Tkachev, Phys. Rev. Lett. 74 (1995) 3093;
D. V. Semikoz and I. I. Tkachev, Phys. Rev. D 55 (1997) 489.
\bibitem{gravcool}
E. Seidel and Wai-Mo Suen, Phys. Rev. Lett. 72 (1994) 2516; 
S. Khlebnikov, astro-ph/9911218.
\bibitem{it87}
I. I. Tkachev, Phys. Lett. B 191 (1987) 41.
\bibitem{ZN}
Ya. B. Zeldovich and I. D. Novikov, {\it Relativistic Astrophysics}, v.1,
the University of Chicago Press, Chicago (1971).
\bibitem{gmm}
A. A. Grib, S. G. Mamaev, and V. M. Mostepanenko, {\it Quantum Effects
in Strong External Fields}, [in Russian], Atomic Energy Press, Moscow (1980).
\bibitem{firma}
C. Firmani et al, astro-ph/0002376.


\end{thebibliography}
\end{document}